\documentclass[twocolumn,showpacs,preprintnumbers,prb,aps,amssymb,superscriptaddress]{revtex4}
\usepackage{graphicx}
\usepackage{dcolumn}
\usepackage{bm} \usepackage{color}
\usepackage{bm}
\begin{document}

\newcommand{\YRS}{YbRh$_2$Si$_2$}
\newcommand{\Co}{CeCoIn$_5$}
\newcommand{\Rh}{CeRhIn$_5$}
\newcommand{\ie}{{\it i.e.}}
\newcommand{\eg}{{\it e.g.}}
\newcommand{\etal}{{\it et al.}}
\newcommand{\p}[1]{\left(#1 \right)}
\newcommand{\Tn}{$T_{\rm N}$}
\newcommand{\Tstar}{$T^\star$}


\title{Wiedemann-Franz law and non-vanishing temperature scale across the field-tuned quantum critical point of YbRh$_2$Si$_2$ }


\author{J.-Ph. Reid}
\altaffiliation[Present address: ] {School of Physics and Astronomy, University of St Andrews, St Andrews KY16 9SS, UK.}
\affiliation{D\'epartement de physique \& RQMP, Universit\'e de Sherbrooke, Sherbrooke, Qu\'ebec, Canada J1K 2R1}

\author{M.~A.~Tanatar}
\affiliation{D\'epartement de physique \& RQMP, Universit\'e de Sherbrooke, Sherbrooke, Qu\'ebec, Canada J1K 2R1}
\affiliation {Ames Laboratory and Department of Physics and Astronomy, Iowa State University, Ames, Iowa 50011, USA.}

\author{R. Daou}
\altaffiliation[Present address: ] {Laboratoire CRISMAT, CNRS, Caen, France.}
\affiliation{D\'epartement de physique \& RQMP, Universit\'e de Sherbrooke, Sherbrooke, Qu\'ebec, Canada J1K 2R1}

\author{Rongwei Hu}
\altaffiliation[Present address: ] {Rutgers Center for Emergent Materials and Department of Physics and Astronomy, Rutgers University, Piscataway, New Jersey 08854, USA}
\affiliation{Department of Physics, Brookhaven National Laboratory, Upton, New York 11973, USA}

\author{C.~Petrovic}
\affiliation{Department of Physics, Brookhaven National Laboratory, Upton, New York 11973, USA}
\affiliation{Canadian Institute for Advanced Research, Toronto, Ontario, Canada M5G 1Z8}

\author{Louis Taillefer}
\email{Louis.Taillefer@USherbrooke.ca}
\affiliation{D\'epartement de physique \& RQMP, Universit\'e de Sherbrooke, Sherbrooke, Qu\'ebec, Canada J1K 2R1}
\affiliation{Canadian Institute for Advanced Research, Toronto, Ontario, Canada M5G 1Z8}

\date{\today}


\begin{abstract}

The in-plane thermal conductivity $\kappa$ and electrical resistivity $\rho$ of the heavy-fermion metal YbRh$_2$Si$_2$ 
were measured down to 50~mK for magnetic fields $H$ parallel and perpendicular to the tetragonal $c$ axis, 
through the field-tuned quantum critical point, $H_c$, at which antiferromagnetic order ends. 
The thermal and electrical resistivities, $w \equiv L_0 T/\kappa$  and $\rho$, show a linear temperature dependence below 1~K,
typical of the non-Fermi liquid behavior found near antiferromagnetic quantum critical points,
but this dependence does not persist down to $T=0$.
Below a characteristic temperature $T^\star \simeq 0.35$~K, which \textcolor{black}{depends weakly} on $H$,
$w(T)$ and $\rho(T)$ both deviate downward and converge as $T \to 0$.
\textcolor{black}{We propose that $T^\star$ marks the onset of short-range magnetic correlations,
persisting beyond $H_c$.
%
By comparing samples of different purity, we conclude that the Wiedemann-Franz law holds in \YRS, 
even at $H_c$, implying that no fundamental
breakdown of quasiparticle behavior occurs in this material.}
The overall phenomenology of heat and charge transport in \YRS~is similar to that observed  
in the heavy-fermion metal CeCoIn$_5$, near its own field-tuned quantum critical point. 

\end{abstract}






\maketitle


\section{ Introduction }

Quantum criticality has emerged as a central paradigm in the physics of heavy-fermion materials.\cite{Mathur,ColemanSchofield}
Spin fluctuations near a magnetic quantum critical point (QCP) lead to unusual electronic properties, 
deviating from those expected in the standard Fermi-liquid theory of metals. 
These deviations, called ``non-Fermi-liquid (NFL) behavior", include a linear temperature dependence of the electrical resistivity $\rho$, 
in contrast to the expected  $T^2$ behavior, 
and a logarithmic divergence of the specific heat $\gamma \equiv C/T$, as opposed to a constant $\gamma$, as $T \to 0$.
A more profound form of NFL behavior would be a violation of the Wiedemann-Franz (WF) law,\cite{Senthil2004,Coleman2005,Podolsky2007,Kim2009,Senthil2012,Mahajan2013}
a robust property of charged fermions.
This law states that the ratio of the thermal conductivity $\kappa$ of a metal to its electrical conductivity $\sigma = 1 / \rho$ has a universal value in the $T = 0$ limit : 
\begin{equation}
 \frac{\kappa}{\sigma T} = L_0 ~~~~,
 \end{equation}
 where $L_0 \equiv (\pi^2/3) (k_{\rm B}/e)^2$.
Defining the thermal resistivity as $~w\equiv~L_0 T/\kappa$, the WF law may be written as $w = \rho$ at $T=0$,
or $\delta(0) = 0$, where $\delta(T) \equiv w(T) - \rho(T)$ is the difference between heat and charge resistivities.

In the heavy-fermion metal CeCoIn$_5$, a QCP is reached by tuning the magnetic field $H$ to $H_c = 5.3$~T ($H \parallel c$).\cite{Paglione2003,Bianchi2003}
At $H_c$, $w(T)$ and  $\rho(T)$  exhibit a linear $T$ dependence at low $T$, 
and so does $\delta(T)$.\cite{Paglione2006,Tanatar2007}
For currents in the basal plane of the tetragonal structure  ($J \perp c$), $\delta(T)$ deviates downward from its linear $T$ dependence below 
$T \simeq 0.4$~K,
and $\delta(T) \to 0$ as $T \to 0$.\cite{Paglione2006}
For currents normal to the basal plane ($J \parallel c$), however, the linear-$T$ dependence of $\delta(T)$ persists down to the lowest temperatures ($\sim 50$~mK),
and $\delta(T)$ extrapolates to a finite value at $T=0$.\cite{Tanatar2007}
In other words, CeCoIn$_5$ exhibits an anisotropic violation of the WF law.

It is of interest to investigate the WF law in other quantum critical systems.
The heavy-fermion metal YbRh$_2$Si$_2$ is an ideal candidate for such a study.
In zero field, it orders antiferromagnetically below a Neel temperature $T_{\rm N} \simeq 70$~mK, and 
a small magnetic field suppresses $T_{\rm N}$ to zero, producing a field-tuned QCP at $H_c = 0.66$~T for $H\parallel c$,
and at $H_c = 0.06$~T for $H\perp c$,\cite{Gegenwart2002,Custers2003} 
where $c$ is the [001] direction of the tetragonal crystal structure.
NFL behavior is observed in \YRS, for example, as a linear $T$ dependence of the resistivity $(\rho \propto T)$
and a logarithmic $T$ dependence of the specific heat ($C/T \propto \ln T$), for $H$ near $H_c$.\cite{Gegenwart2002,Custers2003}
It was suggested that local critical fluctuations \cite{Si2001} in this material make the entire Fermi surface ``hot" and cause a breakdown of quasiparticles,\cite{Custers2003} 
which could produce a violation of the WF law.

Two recent reports provide conflicting interpretations on the validity of the WF law in YbRh$_2$Si$_2$.\cite{Pfau2012,Machida2012}
The data by Pfau {\it et al.},\cite{Pfau2012} 
with $H \perp c$, show that, at $H_c$, $w \simeq \rho$ at the lowest measured temperature ($\sim 30$~mK).
However, the authors argue that a contribution to heat transport from paramagnons must be subtracted from the measured $\kappa$,
and this implies that the purely electronic $\delta(T)$ remains finite as $T \to 0$, so that the WF law is violated at $H_c$.
The data by Machida {\it et al.},\cite{Machida2012} 
with $H \parallel c$, show that, at $H_c$, $w(T) \to \rho(T)$ as $T \to 0$.
Here, the authors argue that the WF law is in fact satisfied at $H_c$.

In this Article, we report measurements of the electrical resistivity and thermal conductivity of YbRh$_2$Si$_2$,
performed on high-quality single crystals for both field orientations.
\textcolor{black}{
Both $w(T)$ and $\rho(T)$ exhibit a linear temperature dependence below 1~K,
but this dependence does not persist down to 50~mK.
Even at the critical field $H_c$, it ends
at a temperature $T^\star \simeq 0.35$~K.
Below $T^\star$, $w(T)$ and $\rho(T)$ start converging so
as to satisfy the WF law at $T \to 0$.
We argue that the WF law is universally obeyed in \YRS~ -- below, at and above
the field-tuned quantum critical point $H_c$, for both $H\perp c$ and $H\parallel c$.
}
Because a similar, albeit sharper, drop in (and convergence of) $w(T)$ and $\rho(T)$ occurs below the 
antiferromagnetic ordering temperature $T_{\rm N}$ in \YRS~at $H=0$ and in the antiferromagnetic 
heavy-fermion metal CeRhIn$_5$,\cite{Paglione2005} 
we propose that $T^\star$ marks the onset of short-range magnetic correlations.
As we shall show, the overall behavior of in-plane transport in \YRS~is similar to that of in-plane transport in CeCoIn$_5$.

\begin{table}[b]
\caption{Characteristics of the YbRh$_2$Si$_2$ samples used to test the Wiedemann-Franz law. 
The residual resistivity ratio is defined as the ratio of resistance at room temperature (300~K) to resistance 
extrapolated to $T=0$ ($\rho_0$): 
$RRR \equiv \rho(300~{\rm K})/\rho_0$.
To remove the uncertainty on $\rho_0$ that comes from the geometric factor, 
we set $\rho(300\text{K})$ = $80~\mu\Omega$~cm.}
\vspace{.2cm}
\begin{tabular}{lcc}
\hline
Sample   &	RRR   &	$\rho_0$      	\\ 
\hline
Pfau {\it et al}.\cite{Pfau2012}~~no.~1        & 			50		&		1.6 		\\
Pfau {\it et al}.\cite{Pfau2012}~~no.~2        & 			73		&		1.1 		\\
Machida {\it et al}.\cite{Machida2012}       &       90                     &			0.9	\\
A      &				105     &			0.75 	\\
B        &				120        & 		0.66	   \\
\hline
\label{samples}
\end{tabular}
\end{table}

\section{Experimental}

\begin{figure}
\centering
\includegraphics[width=8cm]{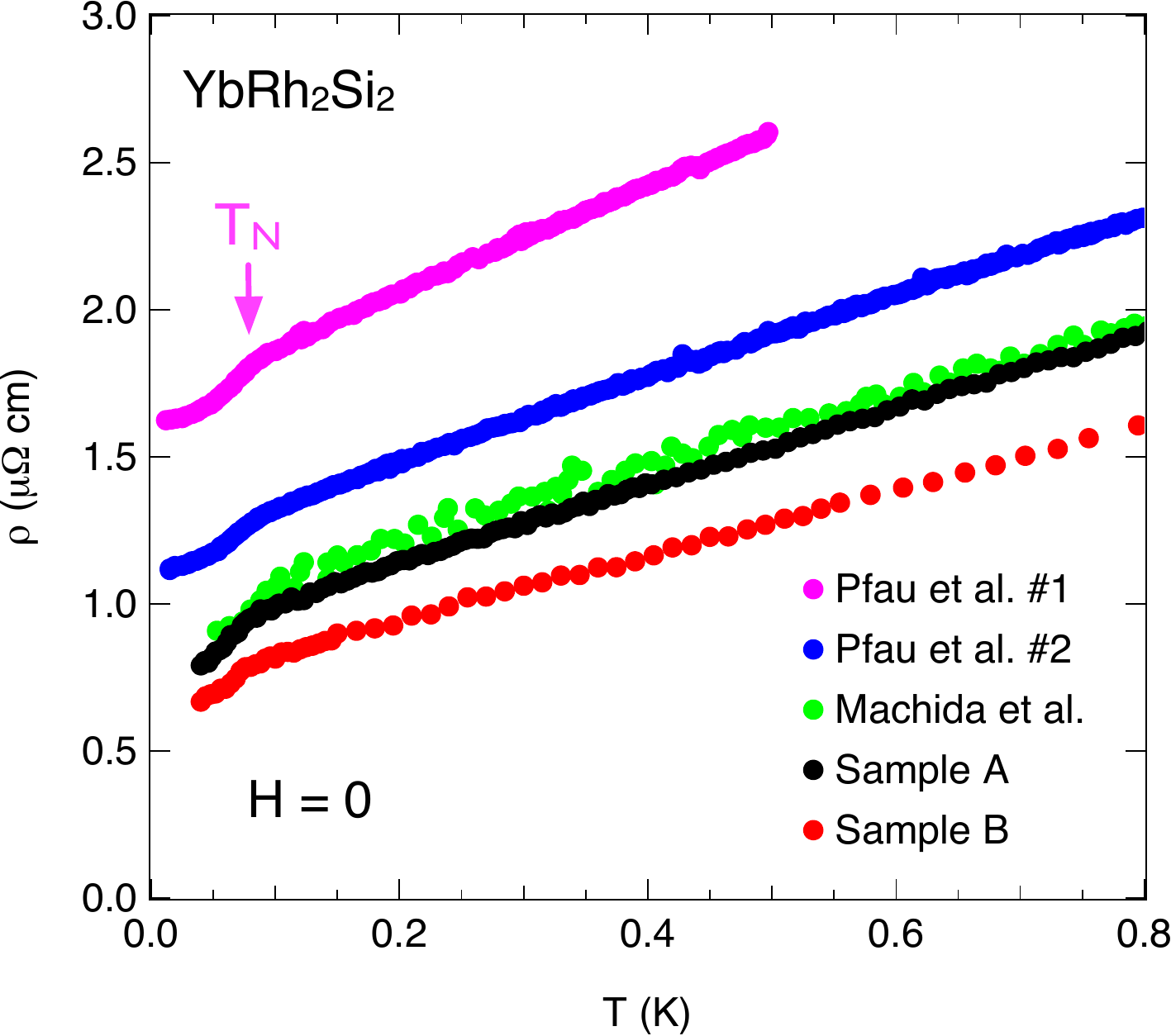}
\caption{Temperature dependence of the in-plane electrical resistivity of five high-purity single crystals of YbRh$_2$Si$_2$, in zero field. 
Data for the two samples used in this study (sample A, black; sample B, red) are compared to previous data, 
from Pfau {\it et al.}\cite{Pfau2012} (no.~1, cyan; no.~2, blue) 
and  Machida {\it et al.}\cite{Machida2012} (green). 
The onset of antiferromagnetic order at $T_{\rm N}$ (arrow) is seen to cause a distinct drop in $\rho(T)$.
}
\label{characterization}	
\end{figure}

High-quality single crystals of YbRh$_2$Si$_2$ were grown by the In flux method with a molar ratio of YbRh$_2$Si$_2$:In = 5:95.
Our method is similar to that of ref.~\onlinecite{Trovarelli2000}, 
but without the use of tantalum tubes. 
Starting ingredients were mixed in an alumina crucible and sealed into a quartz tube. 
The quartz tube was heated to 1150~C, held constant for 2 hours and then cooled to 800~C where crystals were decanted.
The platelet crystals had dimensions up to 3$\times$3$\times$0.2 mm$^3$ and were of high purity, 
as confirmed by their high residual resistivity ratio $RRR$.
Two samples, labelled A and B, were cut for electrical resistivity and thermal conductivity measurements, 
with their long side parallel to the [100] crystallographic direction ($a$ axis), 
for a length of 1.5-2~mm and cross-section of 0.1~mm $\times$ 0.1~mm. 
Four contacts were made on each sample by soldering silver wires with a silver-based alloy, giving
a contact resistance of 1-2~m$\Omega$ at low temperature. 
The samples have $RRR = 105$ (sample A) and 120 (sample B),
slightly higher than the $RRR$ of crystals used in the two previous studies\cite{Pfau2012,Machida2012} 
of the WF law in \YRS~(see Table~\ref{samples}),
but slightly lower than the highest value of $\sim 150$ reported so far.\cite{Custers2003,Gegenwart2006,Knebel2006} 
In Fig.~\ref{characterization}, the zero-field $\rho (T)$ of both samples is compared to the data of refs.~\onlinecite{Pfau2012}
and~\onlinecite{Machida2012}.
All data are in good agreement, modulo a rigid shift due to the different $\rho_0$ values.

Thermal conductivity was measured using the same four contacts as in the four-probe resistivity measurement, in a standard one-heater-two-thermometers technique. \cite{Boaknin2001} 
By using the same contacts, the relative uncertainty between heat and charge transport measurements is removed, and a precise comparison of $w(T)$ and $\rho(T)$ can be made.
For sample A, the magnetic field was applied parallel to the current direction, in the basal plane of the tetragonal structure: $J \parallel a$ and $H \perp c$;
for sample B:  $J \parallel a$ and $H \parallel c$.
%


\begin{figure*}
\centering
\includegraphics[width=17cm]{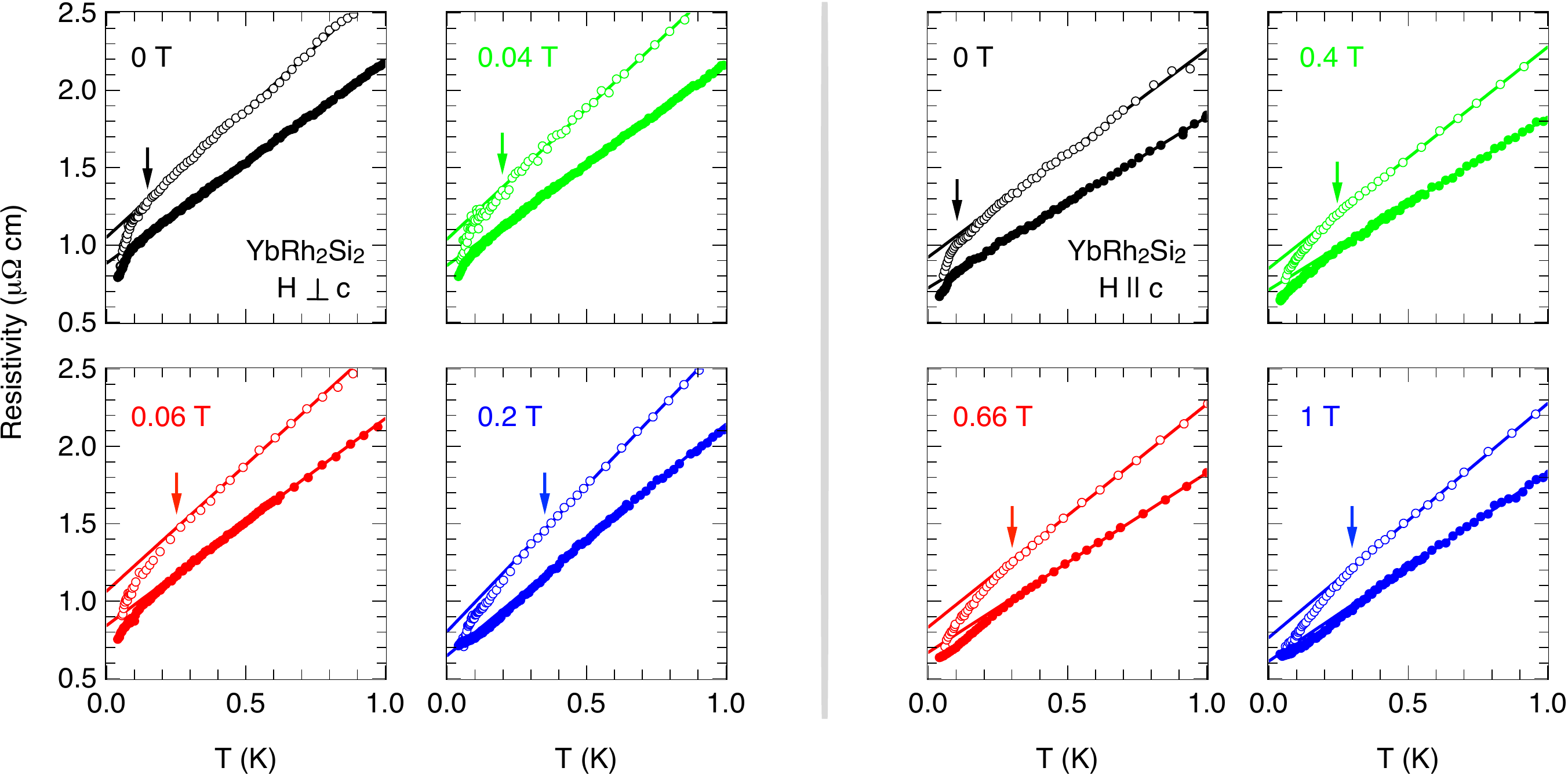}
\caption{
Temperature dependence of the electrical resistivity ($\rho$, closed circles) and thermal resistivity ($w \equiv L_0 T/\kappa$, open circles) of YbRh$_2$Si$_2$,
for currents in the basal plane ($J \parallel a$).
In the four panels on the left, data from sample A were obtained with a magnetic field $H \perp c$, for different field strengths as indicated.
In the four panels on the right, data from sample B were obtained with $H \parallel c$.
Data in red correspond to the field-tuned quantum critical point, at $H_c \simeq 0.06$~T (left;  $H \perp c$) and $H_c \simeq 0.66$~T (right;  $H \parallel c$),
respectively. 
Above a characteristic temperature $T^\star$ (arrow), $w(T)$ and $\rho(T)$ are both linear in temperature; below $T^\star$, they both deviate downward,
and converge as $T \to 0$.  $T^\star$ remains finite even at $H_c$.
}
\label{RhoW}	
\end{figure*}

\begin{figure*}
\centering
\includegraphics[width=17cm]{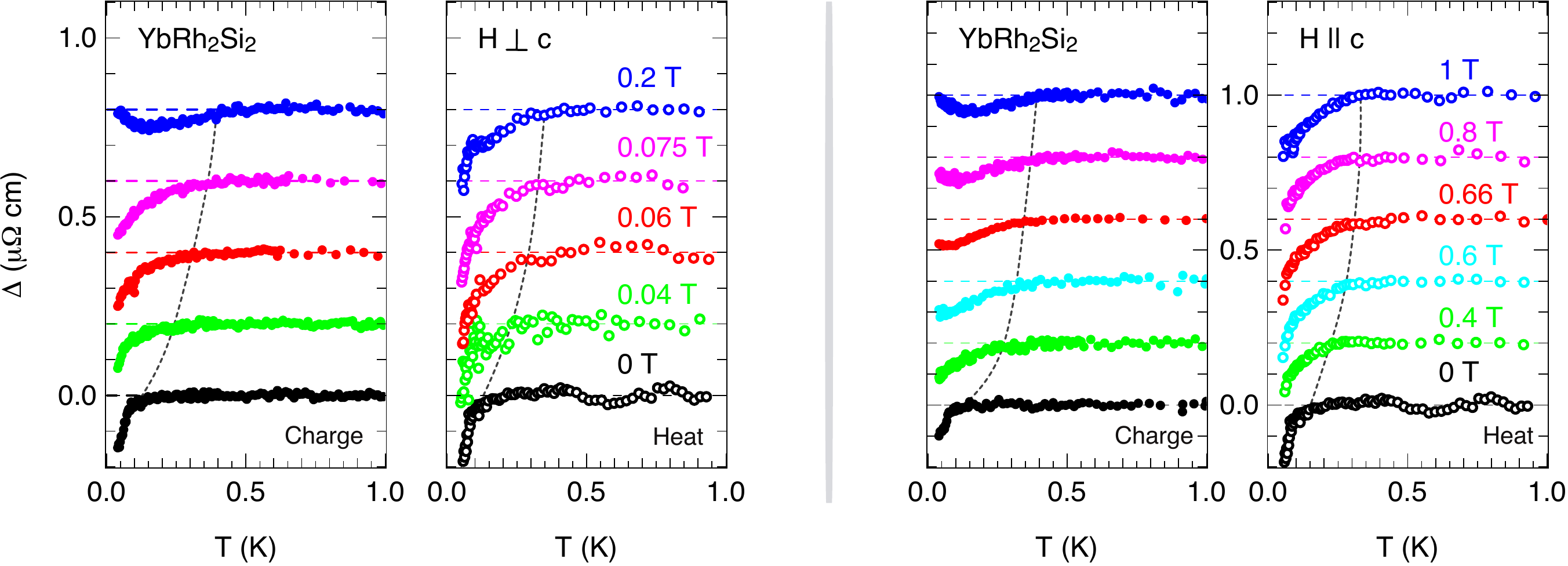}
\caption{
Electrical resistivity $\rho(T)$ and thermal resistivity $w(T)$ as a function of temperature, for field directions and strengths as indicated.
A linear fit has been subtracted from the raw data, so that the quantities being plotted are
$\Delta(T) = \rho_{\text{fit}}-\rho(T)$ (charge; closed circles) and
$\Delta(T) = w_{\text{fit}}-w(T)$ (heat; open circles), 
where $ \rho_{\text{fit}}$ and $w_{\text{fit}}$ are a linear fit to $\rho(T)$ and $w(T)$, respectively,
between $T = T^\star$ and $T = 1.0$~K.
The two panels on the left show data from sample A (with $H \perp c$); the two panels on the right show data  from sample B (with $H \parallel c$).
In all curves, a downward deviation in $\rho(T)$ and $w(T)$ occurs below $T^\star$ (black dashed line).
}
\label{DeltaRhoW}
\end{figure*}

\begin{figure*}
\centering
\includegraphics[width=17cm]{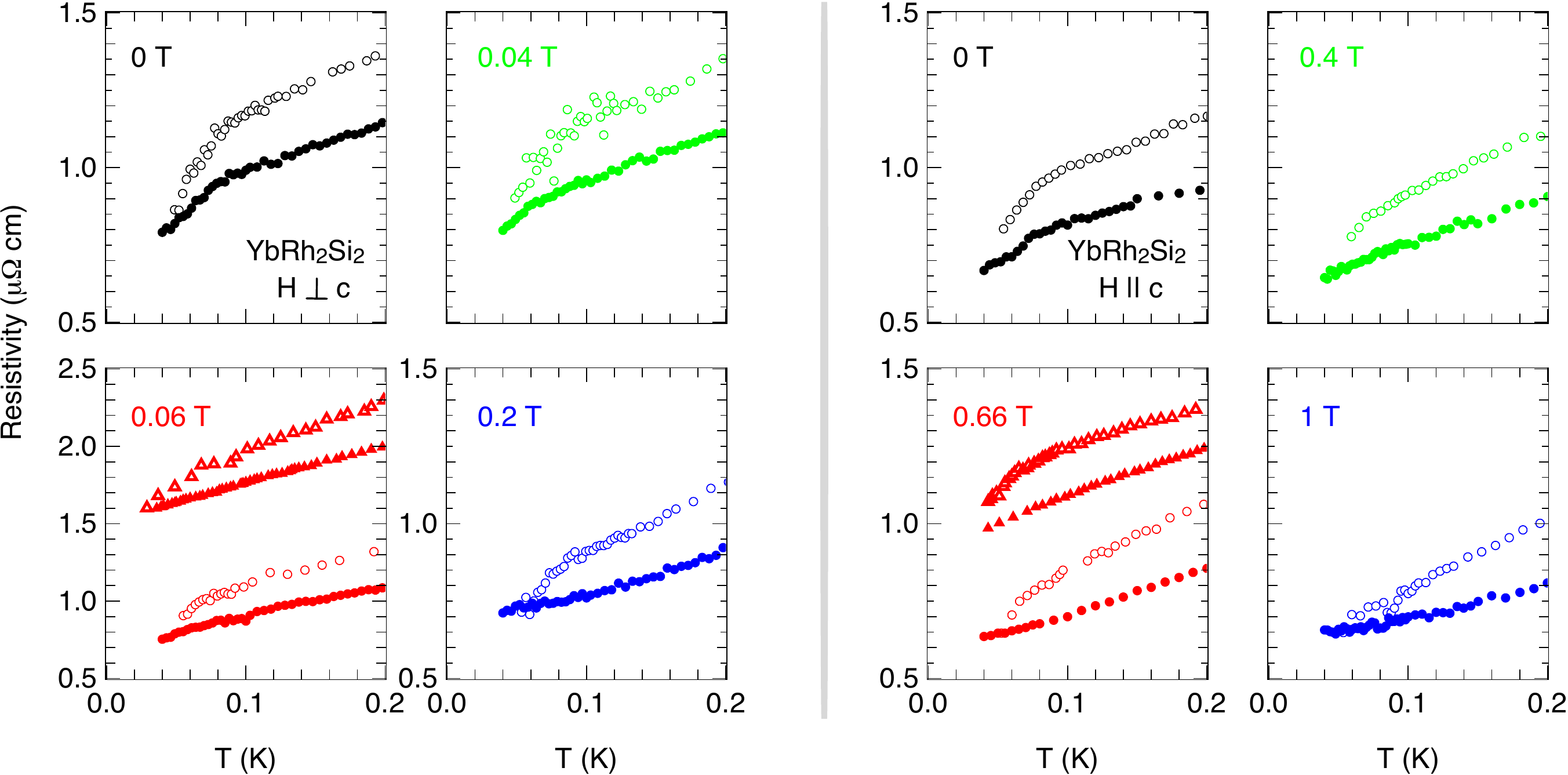}
\caption{
Zoom on the data of Fig.~\ref{RhoW} below 0.2~K.
With decreasing temperature, the thermal resistivity $w(T)$ is seen to drop towards $\rho(T)$ for both field directions and all
field strengths, causing the two resistivities to converge as $T \to 0$.
The raw data of Pfau {\it et al}.\cite{ Pfau2012} (red triangles; $H \perp c$) and Machida {\it et al}.\cite{Machida2012} (red triangles; $H \parallel c$)
are displayed for comparison, at $H = H_c$.
Since $w = \rho$ as $T \to$ 0 in all cases, 
\textcolor{black}{the raw data satisfies the Wiedemann-Franz law} for all values of $H$, in both field directions. 
}
\label{RhoWzoom}	
\end{figure*}

\section{Results}

\begin{figure*}
\centering
\includegraphics[width=17cm]{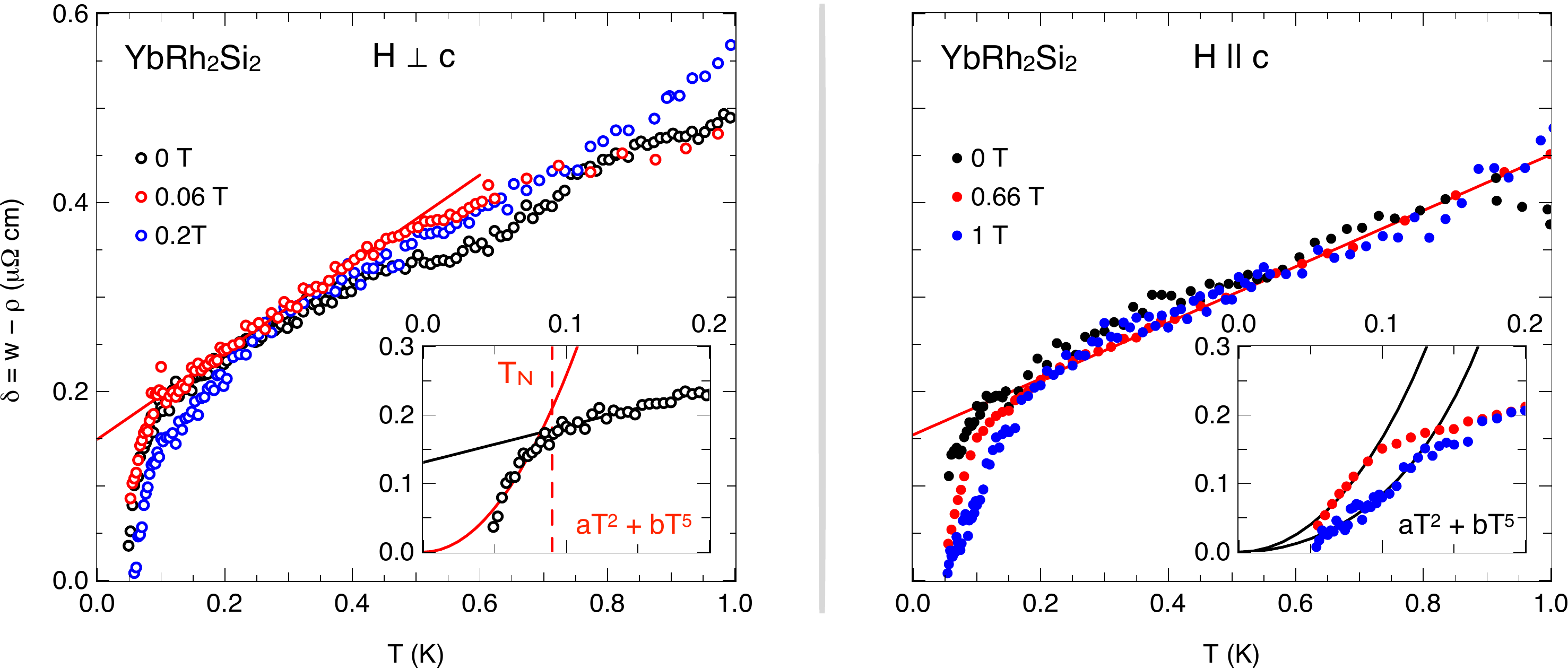}
\caption{
Difference between the thermal and electrical resitivities of YbRh$_2$Si$_2$, $\delta(T) \equiv w(T) - \rho(T)$, as a function of temperature,
for $H \perp c$ (left) and $H \parallel c$ (right). Data are plotted for three values of $H$, as indicated.
{\it Insets}:
Zoom below 0.2~K. 
The vertical dotted line marks \Tn\ (at $H=0$).
Solid lines show the function $\delta(T) = aT^2 + bT^5$, which provides a good description of $\delta(T)$
in the antiferromagnetic heavy-fermion metal CeRhIn$_5$ below $T_{\rm N}$ (see ref.~\onlinecite{Paglione2005}). 
}
\label{delta}
\end{figure*}

Fig.~\ref{RhoW} shows $w(T)$ and $\rho(T)$ below 1~K, for $H \perp c$ and  $H \parallel c$,
at four different values of the applied field.
In all cases, both resistivities show a linear $T$ dependence, a standard signature of NFL behavior, typical of systems close to an antiferromagnetic QCP.
\cite{Doiron-Leyraud2009,Taillefer2010}
But in contrast to the archetypal behavior whereby the linear-$T$ dependence would persist down to $T=0$ at the QCP,\cite{Daou2009}
the linear-$T$ dependence of $w(T)$ and $\rho(T)$ in \YRS~ends at a finite temperature $T^\star$.
In Fig.~\ref{DeltaRhoW}, the downward deviation of $w(T)$ and $\rho(T)$ below $T^\star$ is highlighted by subtracting the linear background.

Below $T^\star$, $w(T)$ drops more rapidly than $\rho(T)$, in such a way that $w(T)$ converges towards $\rho(T)$ as $T \to 0$. 
This is seen most clearly in Fig.~\ref{RhoWzoom}, where we zoom on the raw data at low temperature.
A direct comparison with previously reported data, shown in Fig.~\ref{RhoWzoom} for $H = H_c$,
shows that our data are in good agreement with the data of Pfau {\it et al}.\cite{Pfau2012} for $H \perp c$
and with the data of Machida {\it et al}.\cite{Machida2012} for $H \parallel c$,
\textcolor{black}{modulo the downward shift of our data, due to the higher quality of our samples}.

It is instructive to plot the difference between thermal resistivity and electrical resistivity, $\delta(T) \equiv w(T) - \rho(T)$, 
as done in Fig.~\ref{delta}. 
As discussed previously, \cite{Paglione2006,Tanatar2007,Paglione2005}
$\delta(T)$ reflects the degree to which inelastic scattering is more effective in degrading a heat current than a charge current.
In particular, this includes small-angle scattering processes that change the energy of the carriers without affecting their momentum direction.
As seen in Fig.~\ref{delta}, the $\delta(T)$ curves are essentially the same for $H$ below, at and above $H_c$, for both field directions:
a linear $T$ dependence down to 0.2~K, and then a rapid dive towards zero below 0.2~K.
The dive at $H=0$ is clearly caused by antiferromagnetic order below $T_{\rm N}$ (inset of Fig.~5, left panel).
We propose a related mechanism for the similar dive in $\delta(T)$ at finite $H$, 
namely the onset of magnetic correlations.


\begin{figure}[]
\centering
\includegraphics[width=8cm]{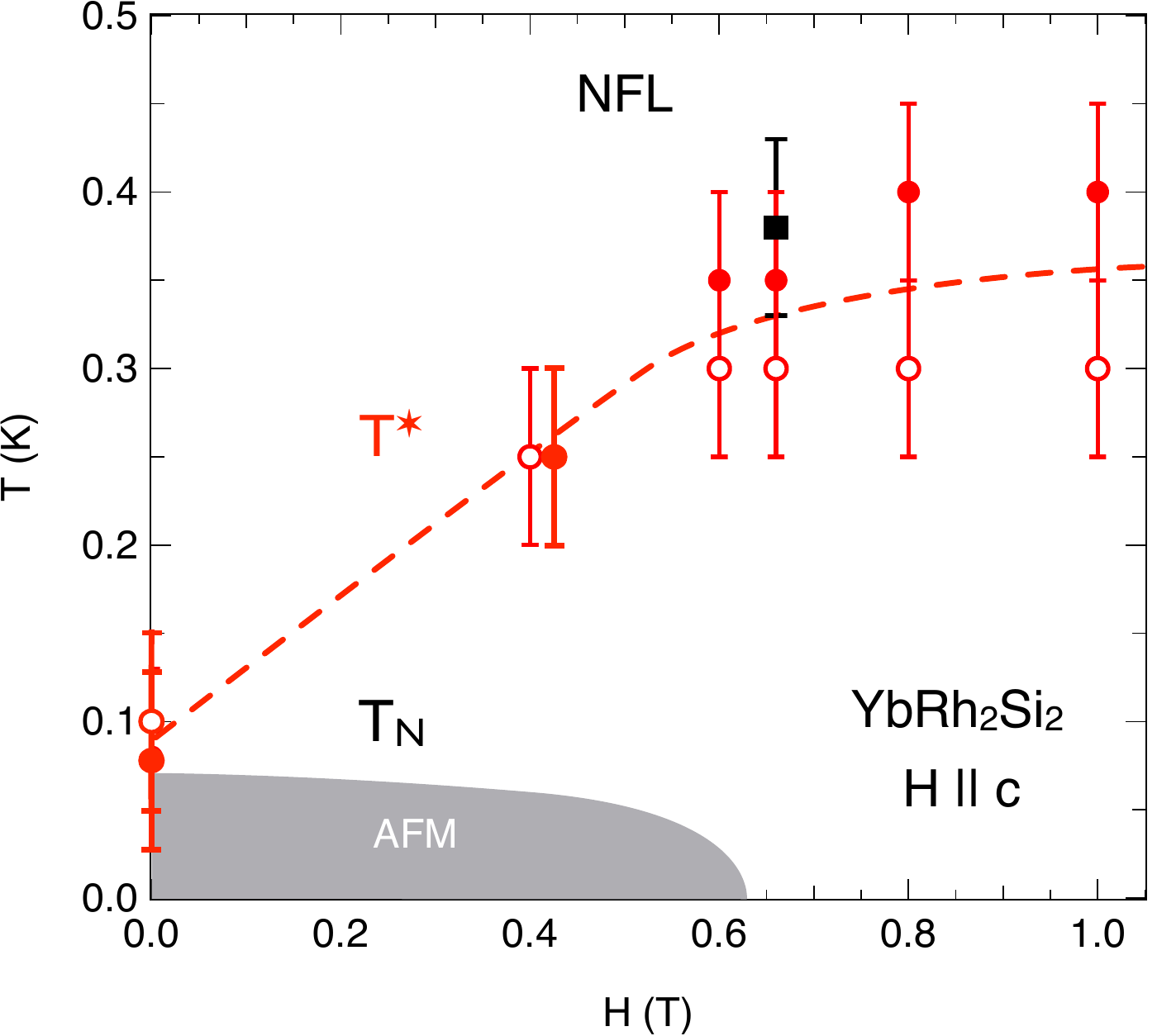}
\caption{
Magnetic field-temperature phase diagram of YbRh$_2$Si$_2$, for $H \parallel c$.
(The phase diagram for $H \perp c$ is very similar.)
The region of long-range antiferromagnetic order is sketched in grey, delineated by the N\'eel temperature $T_{\rm N}$.
The crossover temperature $T^{\star}$ below which the resistivity deviates downward from its linear $T$ dependence
at high temperature is shown for heat transport (open red circles) 
and charge transport (closed red circles).
%
The black square is $T^{\star}$ obtained from \textcolor{black}{resistivity} data in ref.~\onlinecite{Trovarelli2000}, for comparison.
The NFL regime of linear-$T$ resistivity is confined to $T > T^{\star}$, a temperature scale that does not vanish
at the quantum critical point $H_c = 0.66$~T.
We interpret $T^{\star}$ as the onset of magnetic correlations (see text).
}
\label{PhaseDiagram}
\end{figure}

\section{Discussion}

\subsection{Temperature scale $T^\star$}

Our main finding is the existence of a crossover temperature $T^{\star}$ below which $\rho(T)$ and $w(T)$
deviate from their linear-$T$ dependence at higher $T$. 
In Fig.~\ref{PhaseDiagram}, $T^{\star}$  is plotted in the $H$-$T$ diagram of \YRS, for $H \parallel c$.
It traces a line that rises smoothly from $T^{\star} \simeq T_{\rm N}$ at $H=0$ to $T^{\star} \simeq 0.35$~K at $H = H_c$ and beyond.
A very similar line exists in the phase diagram for $H \perp c$.
The presence of such a crossover line frames any description of the electronic behavior in \YRS.

Above the $T^{\star}$ line, the transport properties of \YRS~exhibit the linear-$T$ resistivity 
typical of the NFL behavior observed in the vicinity of a QCP where AF order ends.\cite{Monthoux2007,Taillefer2010}
Indeed, a linear-$T$ electrical resistivity is observed on the border of AF order in the single-band quasi-1D organic
superconductor (TMTSF)$_2$PF$_6$ (ref.~\onlinecite{Doiron-Leyraud2009}) and the multi-band quasi-2D pnictide superconductors
Ba(Fe$_{1-x}$Co$_x$)$_2$As$_2$ (ref.~\onlinecite{Doiron-Leyraud2009}) and BaFe$_2$(As$_{1-x}$P$_x$)$_2$ (ref.~\onlinecite{Kasahara2010}).
It is also observed at the QCP for stripe order -- a pattern of unidirectional charge and spin modulations -- 
in cuprates.\cite{Taillefer2010,Daou2009}
In the heavy-fermion metal CeCoIn$_5$, both $\rho$ and $w$ were shown to display linear-$T$ behavior at the field-tuned QCP.\cite{Tanatar2007}
This QCP is attributed to an underlying AF phase\cite{Pham2006} hidden by the intervening superconductivity.\cite{Paglione2003,Bianchi2003}

In \YRS~at $H=0$,  the two resistivities, and their difference, all drop abruptly below $T_{\rm N} \simeq 80$~mK. 
A sharp drop in $\rho(T)$, $w(T)$ and $\delta(T)$ is also observed in the heavy-fermion material CeRhIn$_5$, 
an antiferromagnet with $T_{\rm N} = 3.8$~K.\cite{Paglione2005}
Clearly, in both materials the scattering is suppressed by the onset of AF order.

Increasing the magnetic field applied to \YRS~causes the onset temperature $T_{\rm N}$ for long-range AF order to go to zero at $H_c$.
However, the temperature scale $T^{\star}$ does not go to zero, but rises instead, to reach a value
at $H_c$ which is roughly 4 times the zero-field value of $T_{\rm N}$ (see Fig.~\ref{PhaseDiagram}). 
Now the resistivity data at $H > 0$ are very similar to those at $H=0$. 
Indeed, the in-field $\rho(T)$ and $w(T)$ drop below $T^{\star}$ in a way that is remarkably similar to the drop
in the zero-field $\rho(T)$ and $w(T)$ below $T_{\rm N}$ (see Figs.~\ref{RhoW} and~\ref{RhoWzoom}). 
The difference $\delta(T)$ also behaves in a  \textcolor{black}{similar} way at $H=0$ and $H >Ê0$ (Fig.~\ref{delta}).
The rapid drop in $\delta(T)$ at low temperature is roughly consistent with the drop seen in CeRhIn$_5$ below its $T_{\rm N}$,\cite{Paglione2005}
which is well described by the function $\delta(T) = a T^2 + bT^5$ (see insets of Fig.~\ref{delta}).
Since the downward deviations in $\rho(T)$, $w(T)$ and $\delta(T)$ for $H=0$ are due to long-range AF order, 
we infer that the similar, but more gradual deviations seen for $H>0$ are due to short-range magnetic order.

Two observations are consistent with short-range magnetic order developing in \YRS~at $H = H_c$ 
below $T^{\star} \simeq 0.35$~K.
The specific heat exhibits an upward deviation from its $\log\p{1/T}$ 
NFL dependence below $T \simeq 0.3$~K,\cite{Oeschler2008,Custers2003}
and the magnetic susceptibility obeys a Curie-Weiss law, with a Curie-Weiss temperature of $-0.32$~K.\cite{Gegenwart2002}
In other words, the presence of short-range order at $H_c$ prevents the NFL behavior in \YRS~from extending down to $T = 0$ at the QCP. 

\textcolor{black}{
Another possible interpretation for the non-vanishing temperature scale $T^\star \simeq 0.3$~K 
is a crossover from a regime of weakly-interacting 2D antiferromagnetic fluctuations to a regime
of strongly-interacting 3D fluctuations.\cite{Abrahams2012}
}

\begin{figure}[t]
\centering
\includegraphics[width=8cm]{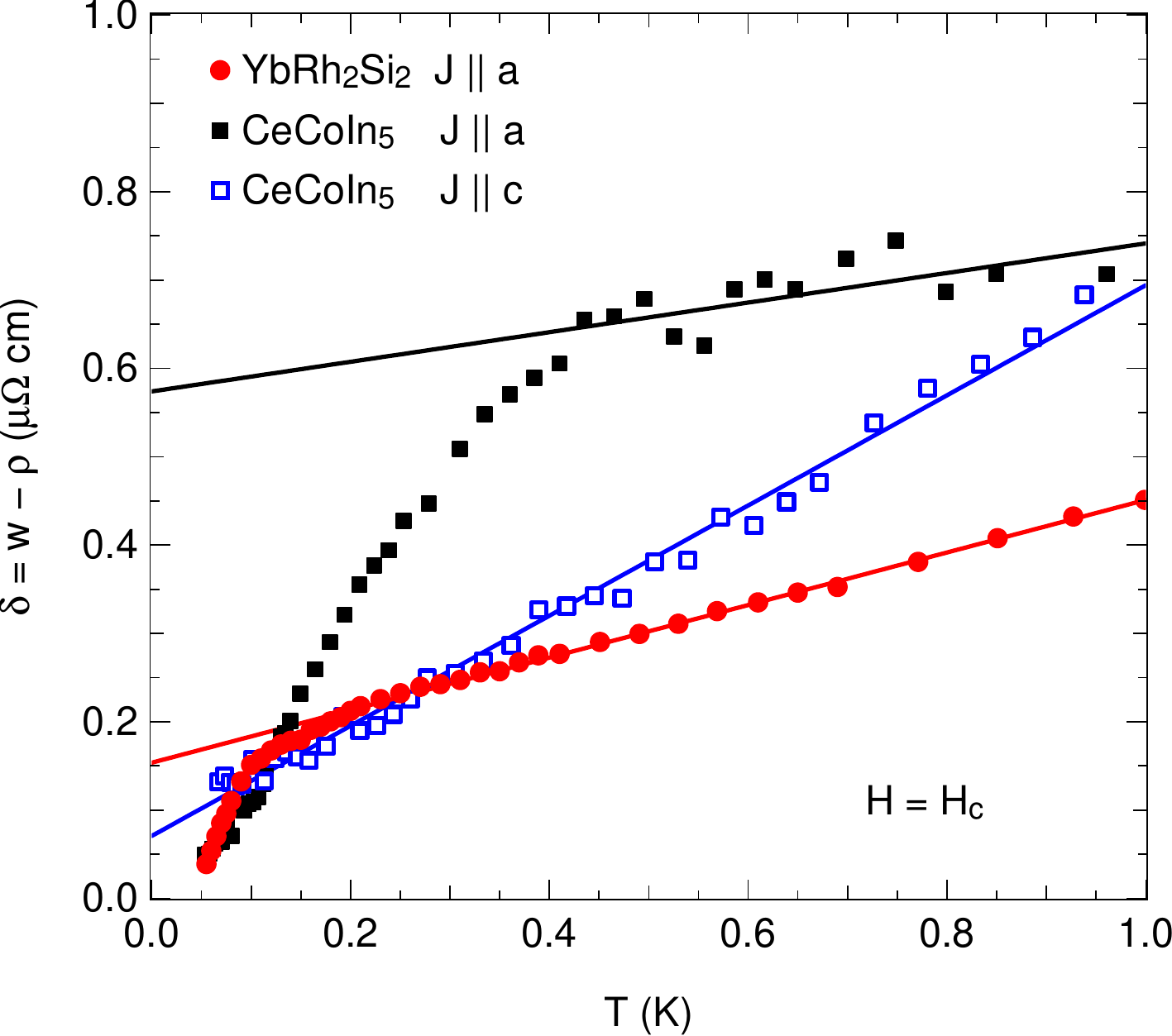}  
\caption{
Temperature dependence of $\delta \equiv w - \rho$, 
the difference between  thermal and electrical resistivities,
for the heavy-fermion metals \YRS~and CeCoIn$_5$, at the critical field $H_c$ of their field-tuned QCP (for $H \parallel c$), 
namely 0.66~T and 5.3~T, respectively.  
The current direction is 
$J\parallel a$ for \YRS~(red circles), 
and either $J\parallel a$ (full black squares)
or
$J\parallel c$ (open blue squares) for CeCoIn$_5$ (data from ref.~\onlinecite{Tanatar2007}).
Solid lines are a linear fit to the high-temperature data. 
For $J \parallel a$, $\delta(T)$ in both materials falls at low temperature, so that the Wiedemann-Franz law is
satisfied, namely $\delta(T) \to 0$ at $T \to 0$. In both cases, the fall occurs well below the temperature scale $T^\star$,
interpreted as the onset of short-range magnetic order.
By contrast, for $J \parallel c$ in CeCoIn$_5$, there is no finite $T^\star$ and $\delta(T)$ is seen to retain
its linear temperature dependence down to the lowest temperature, so that the WF law is violated
in this direction.\cite{Tanatar2007} 
}
\label{comparison}
\end{figure}

\begin{figure}[t]
\centering
\includegraphics[width=8cm]{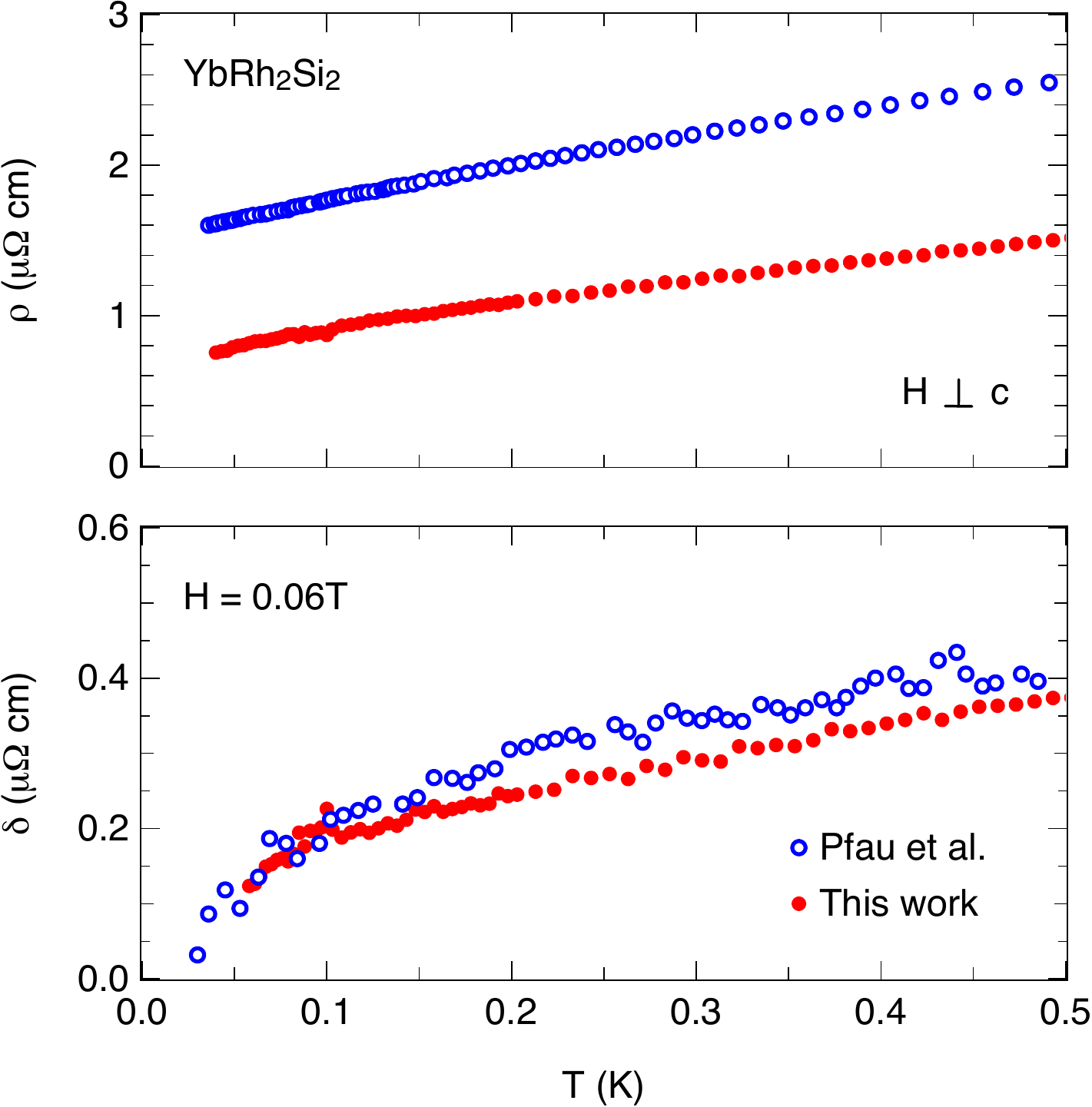}  
\caption{
Comparison of $\rho$ and $\delta$ in two samples of \YRS~with different levels of disorder,
at $H = H_c = 0.06$~T ($H \perp c$).
Our sample A (full red dots) has a $\rho_0$ value 
(Table~I) that is roughly half that of the sample used by Pfau {\it et al}.\cite{Pfau2012}
(open blue circles). 
This factor 2 difference in elastic scattering causes a large rigid shift in $\rho(T)$ at $H_c$ (top panel),
but negligible change, within error bars, in the difference $\delta(T)$ (bottom panel).
This strongly suggests that $\delta(T)$ is entirely due to inelastic scattering, 
and does not contain a residual contribution $\delta(0)$ that persists at $T=0$.
}
\end{figure}

\subsection{Wiedemann-Franz law}

In Fig.~\ref{RhoWzoom}, we saw how $w(T)$ falls at low temperature and converges towards $\rho(T)$, for all $H$.
In Fig.~\ref{delta}, the difference $\delta(T)$ between $w$ and $\rho$ decreases rapidly as $T \to 0$, for all $H$. 
In Fig.~\ref{comparison}, we reproduce $\delta(T)$ at $H = H_c$ for $H \parallel c$.
The linear $T$ dependence of the NFL regime above $T^{\star} \simeq 0.35$~K does not 
persist down to $T=0$ : $\delta(T)$ eventually deviates downwards and exhibits a rapid drop at  low temperature.
In Fig.~\ref{comparison}, we see how $\delta(T)$ in CeCoIn$_5$ and \YRS~are very similar (for $J \parallel a$): 
$\delta(T)$ is linear at high temperature and it drops at low temperature, below $T \simeq 0.4$~K in the former and below $T \simeq 0.1$~K in the latter.
In CeCoIn$_5$, $\delta(T)$ clearly vanishes as $T \to 0$,
showing that $w$ = $\rho$ at $T=0$, satisfying the WF law.
\textcolor{black}{By analogy,} we infer that in the limit of $T=0$ the WF law is also satisfied in \YRS.

In both materials (for $J \parallel a$), the linear $T$ dependence of $\rho(T)$ at $H_c$ is cut off at a finite $T^\star$,
and then, at a temperature well below $T^\star$, $\delta(T)$ starts its rapid drop to zero. 
We associate the recovery of the WF law at $T \to 0$ to the preceding onset of short-range magnetic order at $T^\star$.
%
The scattering then becomes $k$-dependent below $T^\star$ and cold spots appear on the Fermi surface,
causing $\rho(T)$ to deviate from linearity and the WF law to be recovered.

By contrast, when $J \parallel c$, $\rho(T)$ in CeCoIn$_5$ remains perfectly linear down to the lowest measured temperature 
\textcolor{black}{(50~mK)}
and $w(T)$ is also linear all the way down.\cite{Tanatar2007}
So unlike for $J \parallel a$, there is no finite temperature scale, and $T^\star \simeq 0$ (or at least \textcolor{black}{$T^\star < 0.05$~K}).
And for that current direction ($J \parallel c$), $\delta(T)$ retains its linear $T$ dependence all the way down.
This is true not only at $H_c$ (see Fig.~\ref{comparison}),
but also at fields away from $H_c$ (see Fig.~S2 in ref.~\onlinecite{Tanatar2007}).
In that context, the extrapolation to $T=0$ is unambiguous.
Away from the QCP, at $H = 10$~T~$\simeq 2~H_c$, the extrapolation yields $\delta(0) = 0$,
showing the WF law to be cleanly satisfied.
As $H \to H_c$, the entire $\delta(T)$ curves shift rigidly upwards, causing $\delta(0)$ to become non-zero,
rising gradually to reach a maximum value of $\delta(0) \simeq 0.1~\mu\Omega$~cm at $H_c  = 5.3$~T.\cite{Tanatar2007}
Therefore, in CeCoIn$_5$, the violation of the WF law is tuned by the field, cleanly and gradually.
This is not the case in \YRS:
the $\delta(T)$ curves in Fig.~5 at $H <ÊH_c$, $H =ÊH_c$ and $H >ÊH_c$ are not shifted relative to each other,
above $T \simeq 0.2$~K, \ie~in the range from which $\delta(0)$ is extrapolated.
This means that if there really is a non-zero $\delta(0)$ that violates the WF law in \YRS, then it is not tuned by 
the magnetic field (in either orientation), at least in the range up to $H \simeq 2-3~H_c$.

The strong anisotropy of transport in CeCoIn$_5$ confirms that the scattering mechanism in that material -- presumably AF spin fluctuations -- 
is strongly anisotropic, or $k$-dependent.
It would be interesting to see whether the same is true in \YRS, by performing transport measurements for $J \parallel c$.

We stress that our data on \YRS~are in excellent agreement with those of Pfau {\it et al}.\cite{Pfau2012} and Machida {\it et al}.\cite{Machida2012}
(see Fig.~4, for example).
So the conclusion reached by Pfau {\it et al}. that the WFL is violated (by 10~\%) at $H_c$ for $H \perp c$ is not based on a difference in the data,
but rather on their assumption that there is a significant contribution from paramagnons, 
so that $\kappa = \kappa_{\rm electron} + \kappa_{\rm paramagnon}$.
Subtracting this putative contribution ($\kappa_{\rm paramagnon}$) would restore the linear-$T$ dependence of $\delta(T)$ 
so that it would extrapolate to $\delta = 0.15$ at $T\to 0$ (see Fig.~5).
Like Machida {\it et al}., we disagree with this assumption.
One reason is that any paramagnon contribution should presumably decrease as one moves away from the QCP.
However, the drop in $\delta(T)$ does not diminish with increasing field \textcolor{black}{above $H_c$.}
At $H = 3~H_c~(H \perp c)$,  the drop is \textcolor{black}{as pronounced as} at $H_c$ (Fig.~5, left panel).
At $H = 7~H_c~(H \parallel c)$,  the drop in $\delta(T)$ is still very strong.\cite{Machida2012}

\textcolor{black}{
One way to experimentally test whether the WF violation is truly a property of the electrons
in their ground state at $T=0$ is to investigate how the apparent violation depends on the
level of elastic impurity scattering, {\it i.e.} how the violation depends on $\rho_0$.
In Fig.~8, we compare the transport properties of two samples with different levels of disorder,
such that their $\rho_0$ values differ by a factor 2.
We see that $\delta(T)$ is the same at all $T$, within error bars.
This is exactly what one expects for a difference $\delta(T)$ that arises entirely
from inelastic scattering, typically independent of elastic scattering.
In other words, there is no indication of a term in $\delta(T)$ that would persist
at $T=0$, which {\it would} depend on the level of elastic scattering.
}

If the $10\%$ violation claimed by Pfau {\it et al.} were an intrinsic property of the electronic
system in \YRS~at $T=0$, such that the Lorentz ratio $L(0) = 0.9~L_0$, instead of the usual
$L(0) = L_0$, then the curve of $\delta(T)$ for our cleaner sample would have been rigidly shifted
down relative to the $\delta(T)$ curve in their dirtier sample, just as the two $\rho(T)$ curves are
shifted relative to each other.
This is clearly not the case.
We conclude that the weight of evidence is against a violation of the WF law in \YRS.


\section{ Summary and Outlook}

To summarize, we have measured heat and charge transport across the magnetic field-tuned phase diagram 
of \YRS~for fields both parallel and perpendicular to the $c$ axis. 
For a current in the basal plane, the thermal and electrical resistivities exhibit a linear temperature dependence,
characteristic of the non-Fermi-liquid behavior found in the vicinity of an antiferromagnetic quantum critical point.
However, we find that this non-Fermi-liquid behavior does not extend down to $T=0$, even
at the critical field where the long-range AF order in \YRS~vanishes.
The linear-$T$ regime ends at a characteristic temperature $T^\star$, below which the thermal and electrical
resistivities fall and converge as $T \to 0$.
We argue that the Wiedemann-Franz law is satisfied at all fields, even at 
the quantum critical point.
Based on the similarity between transport signatures of $T_{\rm N}$ at $H=0$ and signatures of $T^\star$ at $H > 0$
we infer that $T^\star$ marks the onset of short-range magnetic order.
This short-range order prevents the NFL behavior from persisting down to $T=0$.
The phenomenology is similar to that observed at the field-tuned quantum critical point 
of the heavy-fermion metal CeCoIn$_5$.

The existence of a finite temperature scale $T^\star$ in \YRS~and  CeCoIn$_5$ raises the interesting 
issue of a precursor regime above the onset of long-range antiferromagnetic order, at  $T_{\rm N}$.
An interesting example of this occurs in the iron arsenide BaFe$_2$As$_2$ doped with Co, 
where the resistivity deviates from its linear-$T$ dependence below a temperature $T^\star$ 
that can be as high as 2~$T_{\rm N}$.\cite{Chu2011}

\section*{ ACKNOWLEDGMENTS}

We thank 
J.-P.~Brison,  P.~Coleman, J.~Flouquet, S.~Hartnoll, 
T.~Senthil, and F.~Steglich 
for insightful discussions, and
J.~Corbin for his assistance with the experiments. 
The work at Sherbrooke was supported by the Canadian Institute for Advanced Research and a Canada Research Chair and it was funded by NSERC,
FQRNT and CFI. 
Part of the work was carried out at the Brookhaven National Laboratory, which is operated for the U.S. Department of Energy by Brookhaven Science Associates 
(DE-Ac02-98CH10886) and in the 
Ames laboratory, supported by the U.S. Department of Energy, Office of Basic Energy Sciences, Division of Materials Sciences and Engineering under contract No. DE-AC02-07CH11358.


\end{document}